\documentclass[aps,prd,twocolumn,superscriptaddress,showpacs,nofootinbib]{revtex4-2}

\usepackage{amsmath,amssymb}
\usepackage{graphicx}
\usepackage{booktabs}
\usepackage{multirow}
\usepackage{hyperref,ulem}
\usepackage{xcolor}
\definecolor{dgreen}{RGB}{0,204,0}
\definecolor{dora}{RGB}{255,150,0}
\definecolor{red1}{RGB}{255,0,0}

\newcommand{\Msun}{M_\odot}
\newcommand{\Lam}{\Lambda}
\newcommand{\rhosat}{\rho_0}

\begin{document}

\title{Hyperonic equation of state for neutron stars: \\
A systematic Bayesian comparison of density-dependent and non-linear relativistic mean-field models}

\author{Pedro Sanson}
\email{pedro.sanson@student.uc.pt }
\affiliation{Department of Physics, CFisUC, University of Coimbra, P-3004 - 516 Coimbra, Portugal}

\author{Tuhin Malik}
\email{tm@uc.pt}
\affiliation{Department of Physics, CFisUC, University of Coimbra, P-3004 - 516 Coimbra, Portugal}

\author{Constança Providência}
\email{cp@uc.pt}
\affiliation{Department of Physics, CFisUC, University of Coimbra, P-3004 - 516 Coimbra, Portugal}

\date{\today}

\begin{abstract}
A systematic Bayesian inference study of the equation of state (EOS) of dense matter with strangeness is presented  extending five relativistic mean-field (RMF) models with both constant and density dependent couplings 
to include the full baryon octet. 
The hyperon-nucleon couplings in the scalar channel 
are varied within ranges informed by hypernuclear data, while vector isoscalar couplings are fixed by the SU(6) symmetry quark model. Observational constraints from NICER (PSR~J0030, J0437, J0740) and GW170817, theoretical constraints from chiral effective field theory ($\chi$EFT) and perturbative QCD (pQCD), and experimental constraints from nuclear saturation properties are imposed simultaneously. We find that the inclusion of hyperons systematically reduces the maximum neutron star mass by $0.05$--$0.10\,\Msun$ across all models while increasing the radius at $1.4\,\Msun$ by $0.5$--$0.8$~km. The speed of sound exhibits a characteristic softening at densities $2$--$3\,\rhosat$ coinciding with hyperon onset. All hyperonic models remain consistent with the $2\,\Msun$ constraint. { Models with a more flexible isovector channel span a larger proton fraction when only nucleons are included. However, the extra flexibility is visibly suppressed by hyperons, meaning that the average proton distribution is independent of model flexibility when hyperons are included.  Less flexible models show comparable or slightly increased proton fractions due to EOS stiffening when hyperons are included.} Only a residual number of hyperonic EOS give rise to a mass-radius curve with a negative slope at low masses. A $1.8M_\odot$ neutron star with a radius larger than or similar to the radius of a $1.2M_\odot$ star would give a strong indication that the star contains other baryonic degrees of freedom in addition to nucleons.
\end{abstract}

\maketitle

\section{Introduction}
\label{sec:intro}

Neutron stars (NSs), which are remnants of massive stellar explosions, are among the most extreme objects in the Universe. The equation of state (EOS) of dense matter is a central ingredient in modeling their structure and observable properties such as mass, radius, and tidal deformability~\cite{Malik2022,Typel1999,Char2023}. While purely nucleonic models have been extensively constrained by multi-messenger astrophysical observations, the appearance of heavier baryons -- hyperons ($\Lam, \Sigma, \Xi$) -- inside neutron stars is still debated. The onset of hyperons at densities of $2$--$3$ times nuclear saturation density $\rhosat$ is a robust prediction of relativistic mean-field (RMF) theory~\cite{Glendenning1985,Schaffner1996}.
 
While relativistic mean-field models including hyperons are able to describe two solar mass stars \cite{Weissenborn:2011kb,Bednarek:2011gd,Providencia:2012rx,Lopes:2013cpa,Fortin:2016hny}, microscopic approaches based on many-body descriptions do not describe  massive stars \cite{Vidana:2010ip}. If hyperonic degrees of freedom are included,  maximum masses of 1.5$M_\odot$ or below have been obtained \cite{Schulze:2006vw}, see \cite{Chatterjee:2015pua} for a review.  The inclusion of three-body forces may solve this tension \cite{Lonardoni:2014bwa,Logoteta:2019utx,Gerstung:2020ktv}. The recent calculations carried out with a chiral hyperon-nucleon interaction tuned to the  $p\Lambda$  femtoscopic data from the ALICE Collaboration predicted a maximum mass below 1.4 $M_\odot$, though hyperonic three-body forces were not included \cite{Vidana:2024ngv}.

The so-called ``hyperon puzzle'' refers to the tension between the softening of the EOS caused by hyperons, which tends to reduce the maximum NS mass below $2\,\Msun$, and the observation of massive pulsars such as PSR~J0740+6620 ($M \approx 2.07\,\Msun$)~\cite{Fonseca:2021wxt} and PSR J0348+0432 ($M \approx 2.01\,\Msun$)~\cite{Antoniadis:2013pzd}. The inclusion of repulsive hyperon-hyperon interactions mediated by the hidden-strangeness $\phi$ meson has been shown to stiffen the EOS sufficiently to accommodate massive NSs even with hyperons present~\cite{Bednarek:2011gd,Weissenborn:2011ut}, within a RMF description.

In this work, we perform a systematic Bayesian comparison of five RMF model families --- each in both nucleonic and hyperonic versions --- constrained simultaneously by nuclear saturation properties, $\chi$EFT predictions for pure neutron matter, pQCD at high density, NICER mass-radius measurements, and GW170817 tidal deformability data. Information from hypernuclei \cite{Gal:2016boi} is used to constrain the hyperon-meson couplings. In particular, the couplings determined by reproducing binding energies of hypernuclei within RMF models are taken as reference \cite{Fortin:2017cvt,Provid_ncia_2019,Fortin:2020qin}.

The possibility of using neutron stars to extract information on the hyperon content has  been considered in references \cite{Sun:2022yor,Huang:2024rvj}.  However, the authors concluded that the present uncertainties on the mass and radius of observed neutron stars are still too large and do not allow for drawing any conclusions  on the hyperon content.

In \cite{Malik:2022jqc}, Bayesian inference was performed within an RMF model with density dependent couplings, to determine under which conditions two solar mass neutron stars would contain hyperons. While it was shown that it is possible to build EOS that contain hyperons and predict maximum neutron star masses above 2 $M_\odot$, it was also shown that this requires a quite stiff nuclear EOS, leading to 1.4 $M_\odot$ NS with a radius of $\sim 13$ km, a conclusion similar to the one drawn in \cite{Fortin:2014mya}. An extension of this study to four other RMF formulations, imposing the same constraints on all models, allows us to draw conclusions that are independent of the RMF parametrization/description considered. We hope that the present study will positively contribute to  the discussion of  whether hyperons are present inside neutron stars.

The paper is organized as follows: after a brief introduction of models in Sec. \ref{sec:models}, the results are presented and discussed in Sec. \ref{sec:results} and some conclusions are drawn in Sec. \ref{sec:conclusions}.

\section{The Models}
\label{sec:models}

The EOS of hadronic matter is determined from the Lagrangian density that describes the hadron system. The degrees of freedom included are: the nucleons $N = (p, n)$; the  hyperons $Y = \Lam, \Sigma^{+,0,-}, \Xi^{0,-}$ in the hyperonic case; the meson fields that describe the nuclear interaction, the scalar isoscalar $\sigma$ field, the vector isoscalar $\omega$ field, and the vector isovector $\varrho$ field, with masses $m_i$, $i = \sigma, \omega, \varrho$;  for the hyperonic extension, we additionally include the hidden-strangeness vector meson $\phi$ ($m_\phi = 1020$~MeV), which is important to describe the hyperon-hyperon interaction; in addition, electrons and muons are included for charge neutrality. 

The Lagrangian density is given by
\begin{align}
\mathcal{L} &= \bar\Psi[\gamma^\mu(i\partial_\mu - \Gamma_\omega \omega_\mu - \Gamma_\varrho \, \boldsymbol{t} \cdot \boldsymbol{\varrho}_\mu \nonumber \\
&\quad - \Gamma_\phi \phi_\mu)
-(m - \Gamma_\sigma \sigma - \Gamma_{\sigma^*}\sigma^*)]\Psi \nonumber \\
&+ \tfrac{1}{2}(\partial_\mu\sigma\,\partial^\mu\sigma - m_\sigma^2\sigma^2) \nonumber \\
&+ \tfrac{1}{2} m_\omega^2 \omega_\mu \omega^\mu
- \tfrac{1}{4}F^{(\omega)}_{\mu\nu}F^{(\omega)\mu\nu} \nonumber \\
&+ \tfrac{1}{2} m_\varrho^2 \boldsymbol{\varrho}_\mu \cdot \boldsymbol{\varrho}^\mu
- \tfrac{1}{4}\boldsymbol{F}^{(\varrho)}_{\mu\nu}\cdot\boldsymbol{F}^{(\varrho)\mu\nu} \nonumber \\
&+ \tfrac{1}{2}(\partial_\mu\sigma^*\partial^\mu\sigma^* - m_{\sigma^*}^2{\sigma^*}^2) \nonumber \\
&+ \tfrac{1}{2} m_\phi^2 \phi_\mu \phi^\mu
- \tfrac{1}{4}F^{(\phi)}_{\mu\nu}F^{(\phi)\mu\nu}
+ \mathcal{L}_{\rm NL},
\label{eq:lagrangian}
\end{align}
where the parameters $\Gamma_i$ or $g_i$ (for constant couplings), $i = \sigma, \omega, \varrho, \phi, \sigma^*$ designate the couplings of the mesons to the baryons. For models with constant couplings, the density-dependent couplings $\Gamma_i$ in Eq.~(\ref{eq:lagrangian}) are replaced by couplings $g_i$. The term $\mathcal{L}_{\rm NL}$ is absent for DDH models and includes self-interacting and mixed meson terms in models with constant couplings $g_i$. This contribution to the Lagrangian density is important for defining the density dependence of the EOS when the coupling parameters are constants.

\subsection{Density-dependent models}
\label{sec:ddh}

The density-dependent models include meson--nucleon couplings $\Gamma_i$ that depend on the total baryon density $\rho$:
\begin{equation}
\Gamma_i(\rho) = \Gamma_{i,0}\,h_i(x), \quad x = \rho/n_0, \quad i = \sigma, \omega, \varrho,
\label{eq:dd_coupling}
\end{equation}
with $\Gamma_{i,0}$ the couplings at saturation density $n_0$.
Two parametrizations for $h_i, \, i=\sigma,\, \omega$ are considered:
For the DDB (Malik22) model~\cite{Malik2022}:
\begin{equation}
h_i(x) = \exp[-(x^{a_i} - 1)],
\label{eq:ddb_h}
\end{equation}
and for the DDME (Typel99) model~\cite{Typel1999}:
\begin{equation}
h_i(x) = a_i\,\frac{1 + b_i(x + d_i)^2}{1 + c_i(x + d_i)^2}.
\label{eq:ddh_h}
\end{equation}

The $\varrho$-meson coupling includes an optional $y$ parameter:
\begin{equation}
h_\varrho(x) = y\exp[-a_\varrho(x-1)] + (1-y), \quad 0 < y \leqslant 1.
\label{eq:hy}
\end{equation}
Models with this extra parameter are designated by ``model''$y$
(DDBy, DDMEy). The density dependence of the couplings ensures
that they never increase with density.

\subsection{Non-linear meson terms}
\label{sec:nl}

The NL models employ constant couplings $g_i$ and include nonlinear meson terms in the Lagrangian density:
\begin{align}
\mathcal{L}_{\rm NL} &= -\tfrac{\kappa}{3!}(g_\sigma\sigma)^3
- \tfrac{\lambda_0}{4!}(g_\sigma\sigma)^4 \nonumber \\
&+ \tfrac{\zeta}{4!}(g_\omega^2\omega_\mu\omega^\mu)^2
+ \Lam_w\, g_\varrho^2\,\boldsymbol{\varrho}_\mu\cdot\boldsymbol{\varrho}^\mu\,
g_\omega^2\,\omega_\nu\omega^\nu.
\label{eq:lnl}
\end{align}
The parameters $\kappa$, $\lambda_0$, $\zeta$, and $\Lam_w$ control various aspects of nuclear matter: $\kappa$ and $\lambda_0$ affect the incompressibility at saturation, $\zeta$ modulates the high-density EOS behavior, and the $\omega$--$\varrho$ term determines the density dependence of the symmetry energy.

\subsection{Hyperonic extension}
\label{sec:hyperons}

The full baryon octet ($n, p, \Lam, \Sigma^+, \Sigma^0, \Sigma^-, \Xi^0, \Xi^-$) is included in beta equilibrium with electrons and muons. The hyperon couplings are expressed as ratios relative to the nucleon couplings:
\begin{equation}
g_{iY} = x_{iY} \cdot g_{iN}, \quad  \Gamma_{iY}(\rho) = x_{iY} \cdot \Gamma_{iN}(\rho)
\label{eq:coupling_ratio}
\end{equation}
where $i = \sigma, \omega, \varrho, \phi$ and $Y = \Lam, \Sigma, \Xi$. In the models with density-dependent couplings, the density dependence is the same for hyperons and for nucleons.

The \emph{vector couplings}-isoscalar  are fixed by SU(6) quark-model symmetry:
\begin{align}
x_{\omega\Lam} &= x_{\omega\Sigma} = \tfrac{2}{3}, &
x_{\omega\Xi} &= \tfrac{1}{3}, \nonumber \\
x_{\phi\Lam} &= x_{\phi\Sigma} = -\tfrac{\sqrt{2}}{3}, &
x_{\phi\Xi} &= -\tfrac{2\sqrt{2}}{3}, \nonumber \\
x_{\phi N} &= 0, &
\label{eq:su6}
\end{align}
The $\varrho$-meson coupling ratio $x_{\varrho Y} = 1$ for all $Y$ is the same for all baryons; the isospin projection $I_{3B}$ enters the field equation and defines the strength of the interaction. The \emph{scalar couplings} ($x_{s\Lam}, x_{s\Sigma}, x_{s\Xi}$) are three additional free parameters sampled in the Bayesian inference, with prior ranges informed by hypernuclear data (Table~\ref{tab:priors}) \cite{Fortin:2017cvt,Provid_ncia_2019,Fortin:2020qin}.

In DDH models, the density dependence of the $\phi$ coupling is taken to be the same as that of the $\omega$ nucleon coupling. Although the $\sigma^*$ meson ($m_{\sigma^*} = 975$~MeV), a scalar-isoscalar meson with hidden strangeness, can provide additional attraction among strange baryons~\cite{Schaffner:1993qj}, in the present work all $\sigma^*$--baryon couplings are set to zero ($x_{\sigma^* B} = 0$ for all $B$) in all five models. This choice avoids introducing additional poorly constrained parameters; the effect of a nonzero $\sigma^*$ coupling will be explored in future work. Table~\ref{tab:fixed_couplings} summarizes the fixed coupling ratios from SU(6) symmetry used in this work.

\begin{table}[t]
\caption{Fixed hyperon coupling ratios from SU(6) quark-model symmetry. The $\varrho$-meson coupling ratio $x_\varrho = 1$ for all baryons; the isospin projection $I_{3B}$ enters the field equations.}
\label{tab:fixed_couplings}
\begin{ruledtabular}
\begin{tabular}{lcccc}
Baryon & $x_\omega$ & $x_\phi$ & $x_\varrho$ & $I_3$ \\
\midrule
$p$          & $1$    & $0$             & 1 & $+1/2$ \\
$n$          & $1$    & $0$             & 1 & $-1/2$ \\
$\Lam$       & $2/3$  & $-\sqrt{2}/3$   & 1 & $0$ \\
$\Sigma^+$   & $2/3$  & $-\sqrt{2}/3$   & 1 & $+1$ \\
$\Sigma^0$   & $2/3$  & $-\sqrt{2}/3$   & 1 & $0$ \\
$\Sigma^-$   & $2/3$  & $-\sqrt{2}/3$   & 1 & $-1$ \\
$\Xi^0$      & $1/3$  & $-2\sqrt{2}/3$  & 1 & $+1/2$ \\
$\Xi^-$      & $1/3$  & $-2\sqrt{2}/3$  & 1 & $-1/2$ \\
\end{tabular}
\end{ruledtabular}
\end{table}

\subsection{Equations of motion}
\label{sec:eom}

In the mean-field approximation, the meson field equations for the hyperonic case are:
\begin{align}
m_{\sigma,{\rm eff}}^2 \bar\sigma &= \sum_{B \in N\cup Y} g_{\sigma B}\,\rho^s_B, \label{eq:sigma} \\
  m_{\omega,{\rm eff}}^2 \bar\omega &= \sum_{B \in N\cup Y} g_{\omega B}\,\rho_B, \label{eq:omega} \\
 m_{\varrho,{\rm eff}}^2 \bar\varrho_{03} &= \sum_{B \in N\cup Y} g_{\varrho B}\,I_{3B}\,\rho_B, \label{eq:rho} \\
m_\phi^2 \bar\phi &= \sum_{B \in N\cup Y} g_{\phi B}\,\rho_B, \label{eq:phi} \\
m_{\sigma^*}^2 \bar\sigma^* &= \sum_{B \in N\cup Y} g_{\sigma^* B}\,\rho^s_B, \label{eq:sigmas}
\end{align}
where $\rho_B = k_{FB}^3/(3\pi^2)$ is the baryon number density,
$\rho^s_B$ the scalar density,   $m^2_{i,{\rm eff}}=m^2_i$ for DDH models, and for NL models are given by
\begin{align}
   m_{\sigma,{\rm eff}}^{2}&= m_{\sigma}^{2}+ \frac{\kappa}{2} g_\sigma^3 \sigma+\frac{\lambda}{6} g_\sigma^4 \sigma^{2} \\
   m_{\omega,{\rm eff}}^{2}&= m_{\omega}^{2}+ \frac{\xi}{3!}g_{\omega}^{4}\omega^{2} +2\Lambda_{\omega}g_{\varrho}^{2}g_{\omega}^{2}\varrho^{2}\\
   m_{\varrho,{\rm eff}}^{2}&=m_{\varrho}^{2}+2\Lambda_{\omega}g_{\omega}^{2}g_{\varrho}^{2}\omega^{2}.
\end{align}

The effective baryon mass is $m^*_B = m_B - g_{\sigma B}\bar\sigma - g_{\sigma^* B}\bar\sigma^*$. For DDH models, a rearrangement self-energy $\Sigma_0^R$ arises from the density dependence of the couplings and enters the chemical potential of every baryon.

\section{Bayesian Inference Framework}
\label{sec:inference}

The posterior is sampled using \texttt{PyMultiNest}~\cite{Buchner2014} with 4500 live points. The total log-likelihood combines the following constraints:

\paragraph{Nuclear matter properties (NMP).} Gaussian constraints at saturation --- binding energy $e_0 = -16.0 \pm 0.2$~MeV, incompressibility $K_0 = 230 \pm 40$~MeV, symmetry energy $J_{\rm sym} = 32.5 \pm 2.5$~MeV. Computed from the \emph{nucleonic} EOS.

\paragraph{$\chi$EFT constraints.} Super-Gaussian constraints on pure neutron matter energy density at sub-saturation densities, with 5\% enlargement~\cite{Huth2022}. Computed from the \emph{nucleonic} EOS.

\paragraph{NICER mass-radius.} KDE likelihoods from posterior samples of PSR~J0030+0451~\cite{Riley:2019yda}, PSR~J0437$-$4715~\cite{Choudhury:2024xbk}, and PSR~J0740+6620~\cite{Riley:2021pdl}. Evaluated on the \emph{hyperonic} EOS for hyperonic models.

\paragraph{GW170817.} Tidal deformability likelihood from the binary NS merger~\cite{Abbott:2017vwq,Abbott:2018exr}.

\paragraph{pQCD.} Constraint at $\rho \approx 8.5\,\rhosat$ using the Komoltsev-Kurkela framework~\cite{Komoltsev:2021jzg}.

The key design choice is that NMP and $\chi$EFT use the nucleonic code (hyperons do not change saturation or PNM properties), while all astrophysical constraints use the hyperonic EOS.

\subsection{Prior ranges}
Table~\ref{tab:priors} lists the prior ranges for all models. The nucleonic parameters are identical between the nucleonic and hyperonic versions; the hyperonic models add three scalar coupling ratios.

\begin{table*}[t]
\caption{Prior ranges for all CEDF models evaluated in this study. All priors are flat (uniform) unless otherwise noted. The saturation density $\rho_0$ has a Gaussian prior $\mathcal{N}(0.153, 0.005^{ 2})$~fm$^{-3}$. The last three rows show the hyperon scalar coupling ratios added for ``+Hyp'' models. The $y$ parameter is included only in DDBy and DDMEy variants.}
\label{tab:priors}
\begin{ruledtabular}
\begin{tabular}{lcccccc}
& \multicolumn{2}{c}{DDB / DDBy} & \multicolumn{2}{c}{DDME / DDMEy} & \multicolumn{2}{c}{RMFNL} \\
\cmidrule(lr){2-3} \cmidrule(lr){4-5} \cmidrule(lr){6-7}
Parameter & Min & Max & Min & Max & Min & Max \\
\midrule
$a_s$ / $g_\sigma$          & 0    & 0.3   & 6.5  & 14.0  & 6.5  & 15.5 \\
$a_v$ / $g_\omega$          & 0    & 0.3   & 6.5  & 15.0  & 6.5  & 15.5 \\
$a_r$ / $g_\varrho$         & 0    & 1.3   & 5.0  & 17.0  & 5.5  & 16.5 \\
$g_{s0}$ / $d_s$ / $\kappa$ & 6.5  & 13.5  & 0    & 1.0   & 0.00476 & 0.0857 \\
$g_{v0}$ / $a_s$ / $\lambda_0$ & 7.5 & 14.5 & 1.0  & 2.0   & $-0.05$ & 0.05 \\
$g_{\varrho 0}$ / $d_v$ / $\zeta$ & 5 & 16 & 0    & 2.0   & 0    & 0.04 \\
$a_r$ (DDME) / $\Lam_w$   & ---  & ---   & 0    & 1.5   & 0    & 0.15 \\
$y$ (DDBy/DDMEy only)      & 0    & 1     & 0    & 1     & ---  & ---  \\
\midrule
$x_{s\Lam}$                 & 0.600& 0.625 & 0.600& 0.625 & 0.600& 0.625 \\
$x_{s\Sigma}$               & 0.430& 0.510 & 0.430& 0.510 & 0.430& 0.510 \\
$x_{s\Xi}$                  & 0.295& 0.335 & 0.295& 0.335 & 0.295& 0.335 \\
\midrule
\textbf{$N_{\rm params}$} (nuc/hyp) & \multicolumn{2}{c}{9--10 / 12--13}
                             & \multicolumn{2}{c}{10--11 / 13--14}
                             & \multicolumn{2}{c}{10 / 13} \\
\end{tabular}
\end{ruledtabular}
\end{table*}

\section{Results}
\label{sec:results}

\subsection{EOS, mass-radius, and speed of sound\label{sec:eos_mr_cs2}}

\begin{table}[t]
\caption{Key astrophysical properties: median and 90\% CI.}
\label{tab:astro}
\begin{ruledtabular}
\begin{tabular}{lccc}
Model & $M_{\rm max}$ [$\Msun$] & $R_{1.4}$ [km] & $\Lam_{1.4}$ \\
\midrule
DDB       & $2.08^{+0.11}_{-0.11}$ & $12.32^{+0.47}_{-0.45}$ & $352^{+115}_{-73}$ \\
DDB+Hyp   & $1.99^{+0.09}_{-0.09}$ & $12.84^{+0.37}_{-0.37}$ & $470^{+123}_{-89}$ \\[3pt]
DDBy      & $2.07^{+0.11}_{-0.10}$ & $12.45^{+0.48}_{-0.44}$ & $381^{+125}_{-80}$ \\
DDBy+Hyp  & $1.99^{+0.09}_{-0.09}$ & $12.99^{+0.39}_{-0.39}$ & $507^{+114}_{-107}$ \\[3pt]
DDME     & $2.07^{+0.11}_{-0.09}$ & $12.21^{+0.40}_{-0.39}$ & $308^{+80}_{-70}$ \\
DDME+Hyp & $2.02^{+0.09}_{-0.10}$ & $12.93^{+0.34}_{-0.35}$ & $555^{+151}_{-105}$ \\[3pt]
DDMEy    & $2.06^{+0.09}_{-0.10}$ & $12.38^{+0.44}_{-0.42}$ & $343^{+114}_{-82}$ \\
DDMEy+Hyp& $2.01^{+0.09}_{-0.11}$ & $13.07^{+0.36}_{-0.35}$ & $576^{+138}_{-107}$ \\[3pt]
RMFNL     & $2.03^{+0.11}_{-0.10}$ & $12.21^{+0.39}_{-0.33}$ & $341^{+91}_{-64}$ \\
RMFNL+Hyp & $1.93^{+0.09}_{-0.09}$ & $12.81^{+0.29}_{-0.32}$ & $504^{+127}_{-99}$ \\
\end{tabular}
\end{ruledtabular}
\end{table}

\begin{figure*}[p]
\centering
\includegraphics[width=0.90\textwidth,height=0.85\textheight,keepaspectratio]{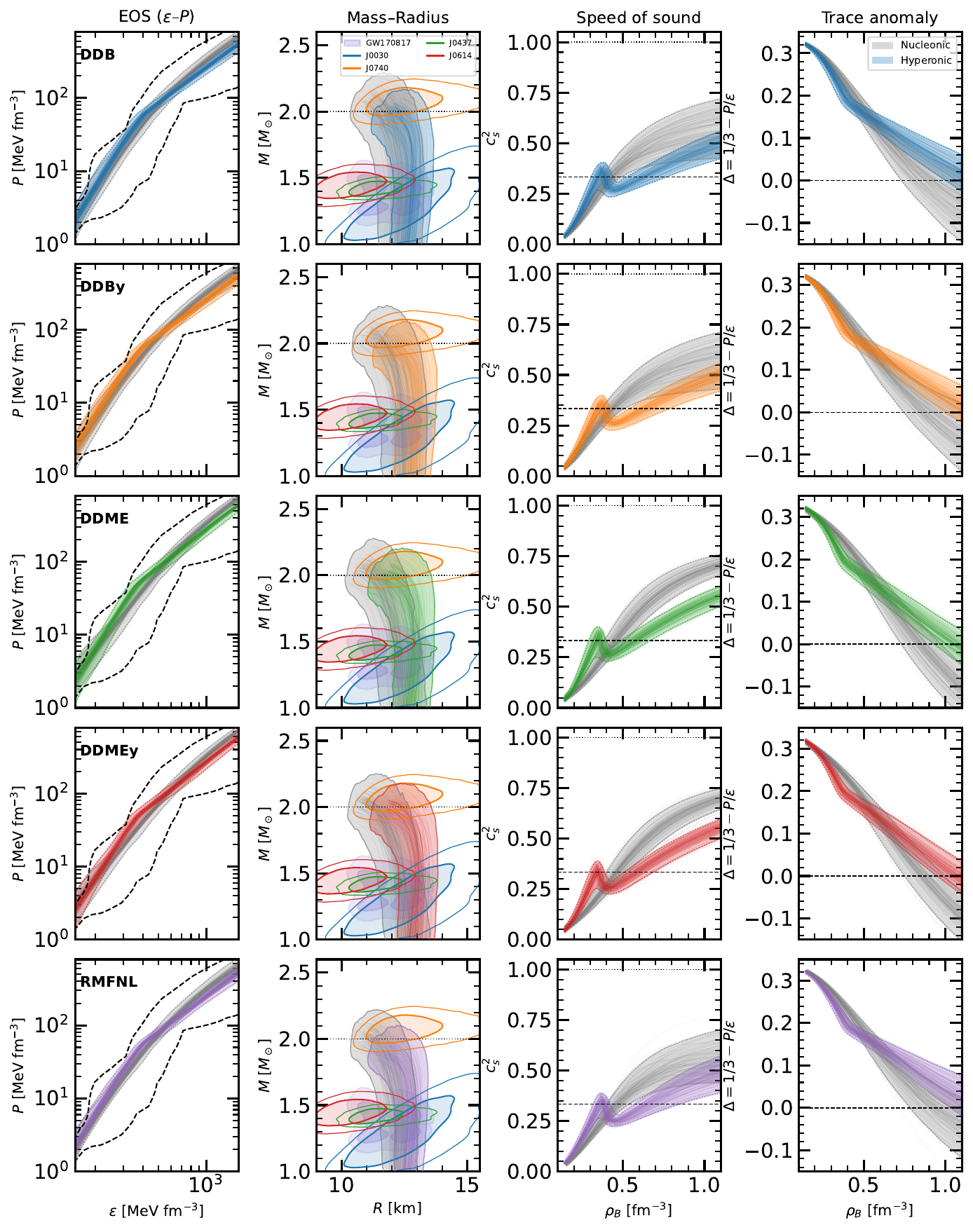}
\caption{Posterior domains for the EOS (log-log), mass-radius (stable branch only), and $c_s^2$  and trace anomaly for all five model families. Gray: nucleonic; colored: hyperonic. Dotted lines: full domain boundaries; thin curves: 150 random posterior samples. M-R panels include NICER and GW170817 constraints. Dashed/dotted lines in $c_s^2$ panels: conformal limit ($1/3$) and causality ($1$).}
\label{fig:5x3}
\end{figure*}

Figure~\ref{fig:5x3} presents the EOS ($\varepsilon$--$P$), mass-radius relation, and speed of sound ($c_s^2$ vs.\ $\rho_B$) for all five model families. In each row, the gray band shows the full nucleonic posterior domain, and the colored band shows the full hyperonic posterior. Several features are robust across all models: (i) The hyperonic EOS is systematically softer at high density; (ii) $M_{\rm max}$ is reduced by $\sim 0.05$--$0.10\,\Msun$ with hyperons, while $R_{1.4}$ increases by $\sim 0.5$--$0.8$~km; (iii) $c_s^2(\rho)$ shows a characteristic dip at $2$--$3\,\rhosat$; (iv) all hyperonic models remain consistent with $M_{\rm max} \gtrsim 2\,\Msun$.

Some key NS  properties obtained for each dataset that sustain the above conclusions  are summarized in Table \ref{tab:astro}, in particular, the median and the 90\% CI for the maximum mass,  radius, and tidal deformability of 1.4$M_\odot$ stars. 

Another interesting characteristic is better observed when comparing the radii of NS with different masses, as shown in Table \ref{tab:radii}: the general trend is that the 1.0$M_\odot$ stars have a larger (smaller) radius than the 1.4$M_\odot$ stars for nucleonic (hyperonic) stars. When comparing  1.0$M_\odot$ with 1.8$M_\odot$ stars, the median radius of the hyperonic star with the more massive star is still larger, except for the RMFNL (Hyp) dataset. Although the difference is not larger than $\sim 100$~m, for the nucleonic stars, the $1.8\,M_\odot$ star may have a radius that is $\sim 200$~m to $700$~m smaller. A massive star with a radius larger than that of a low mass star could indicate the presence of non-nucleonic matter inside the neutron star. A similar conclusion was drawn in \cite{Ferreira:2024hxc,Albino:2025puc}.

\begin{table*}                                                                                   
  \caption{Median and 68\% credible intervals for neutron star radii (in km) at selected masses.}
  \label{tab:radii}                                                            \begin{ruledtabular}                                      
  \begin{tabular}{lccccc}                                                      Model & $R_{1.0}$ & $R_{1.2}$ & $R_{1.4}$ & $R_{1.8}$ & $R_{2.0}$ \\
  \hline                                                                       DDB & $12.26^{+0.27}_{-0.27}$ & $12.31^{+0.27}_{-0.28}$ & $12.31^{+0.28}_{-0.29}$ & $12.09^{+0.32}_{-0.33}$ & $11.70^{+0.39}_{-0.41}$ \\
  DDB (Hyp) & $12.66^{+0.21}_{-0.21}$ & $12.77^{+0.21}_{-0.22}$ & $12.84^{+0.22}_{-0.23}$ & $12.75^{+0.29}_{-0.33}$ & $12.37^{+0.38}_{-0.38}$ \\         
  DDBy & $12.46^{+0.27}_{-0.29}$ & $12.48^{+0.28}_{-0.29}$ & $12.45^{+0.30}_{-0.29}$ & $12.17^{+0.34}_{-0.33}$ & $11.73^{+0.42}_{-0.42}$ \\              
  DDBy (Hyp) & $12.83^{+0.21}_{-0.23}$ & $12.92^{+0.21}_{-0.23}$ & $12.99^{+0.22}_{-0.23}$ & $12.88^{+0.29}_{-0.33}$ & $12.47^{+0.39}_{-0.38}$ \\        
  DDME & $12.35^{+0.26}_{-0.24}$ & $12.29^{+0.24}_{-0.23}$ & $12.21^{+0.23}_{-0.23}$ & $11.80^{+0.27}_{-0.27}$ & $11.30^{+0.38}_{-0.35}$ \\             
  DDME (Hyp) & $12.74^{+0.21}_{-0.20}$ & $12.85^{+0.20}_{-0.19}$ & $12.93^{+0.20}_{-0.20}$ & $12.82^{+0.27}_{-0.33}$ & $12.36^{+0.38}_{-0.45}$ \\       
  DDMEy & $12.57^{+0.29}_{-0.26}$ & $12.49^{+0.28}_{-0.24}$ & $12.37^{+0.26}_{-0.24}$ & $11.89^{+0.27}_{-0.28}$ & $11.32^{+0.37}_{-0.34}$ \\            
  DDMEy (Hyp) & $12.91^{+0.22}_{-0.21}$ & $13.00^{+0.22}_{-0.21}$ & $13.07^{+0.22}_{-0.21}$ & $12.92^{+0.27}_{-0.34}$ & $12.39^{+0.39}_{-0.42}$ \\      
  RMFNL & $12.23^{+0.21}_{-0.19}$ & $12.24^{+0.22}_{-0.20}$ & $12.21^{+0.23}_{-0.21}$ & $11.87^{+0.29}_{-0.30}$ & $11.43^{+0.38}_{-0.36}$ \\             
  RMFNL (Hyp) & $12.66^{+0.17}_{-0.18}$ & $12.75^{+0.17}_{-0.19}$ & $12.81^{+0.18}_{-0.20}$ & $12.57^{+0.30}_{-0.36}$ & $12.22^{+0.33}_{-0.31}$ \\       
  \end{tabular}                                                              
  \end{ruledtabular}
  \end{table*}

\subsection{Particle fractions}
\begin{figure*}[t]
\centering
\includegraphics[width=0.95\textwidth]{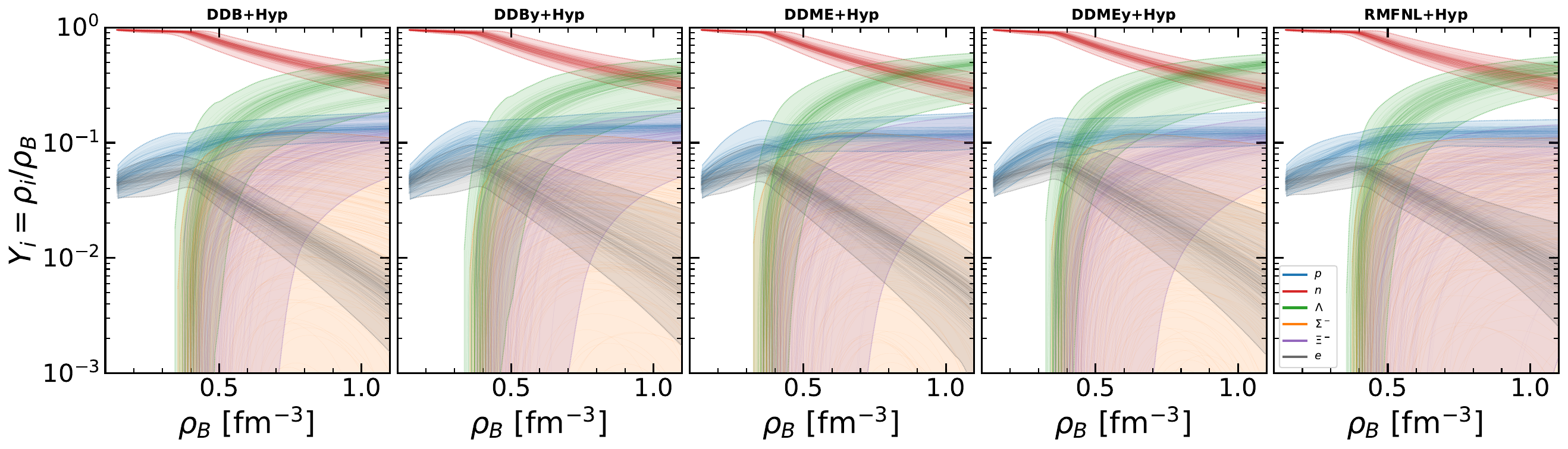}
\caption{Particle fractions $Y_i = \rho_i/\rho_B$ vs.\ baryon density for all five hyperonic models. Species: proton (blue), neutron (red), $\Lam$ (green), $\Sigma^-$ (orange), $\Xi^-$ (purple), electron (gray). Dotted: full domain; thin curves: 200 random samples.}
\label{fig:fractions}
\end{figure*}

Figure~\ref{fig:fractions} shows the particle composition as a function of baryon density. The $\Lam$ hyperon appears first at $\rho \sim 2$--$3\,\rhosat$, followed by $\Sigma^-$ and $\Xi^-$. The logarithmic scale reveals the onset and growth of each species across the posterior. Some general conclusions can be drawn: i) the fraction of $\Lambda$-hyperons becomes larger than that of protons for $\rho_B\gtrsim 0.5$fm$^{-3}$, and that of neutrons for $\rho_B\gtrsim 0.8-1.0$~fm$^{-3}$; ii) there is a competition between the $\Sigma^-$ and the $\Xi^-$ hyperons at onset density,  but at high density, the fraction of $\Xi^-$ is larger; the fraction of electrons is below 1\% for  $\rho_B\sim 1.0$fm$^{-3}$.

\begin{figure*}[t]
\centering
\includegraphics[width=0.95\textwidth]{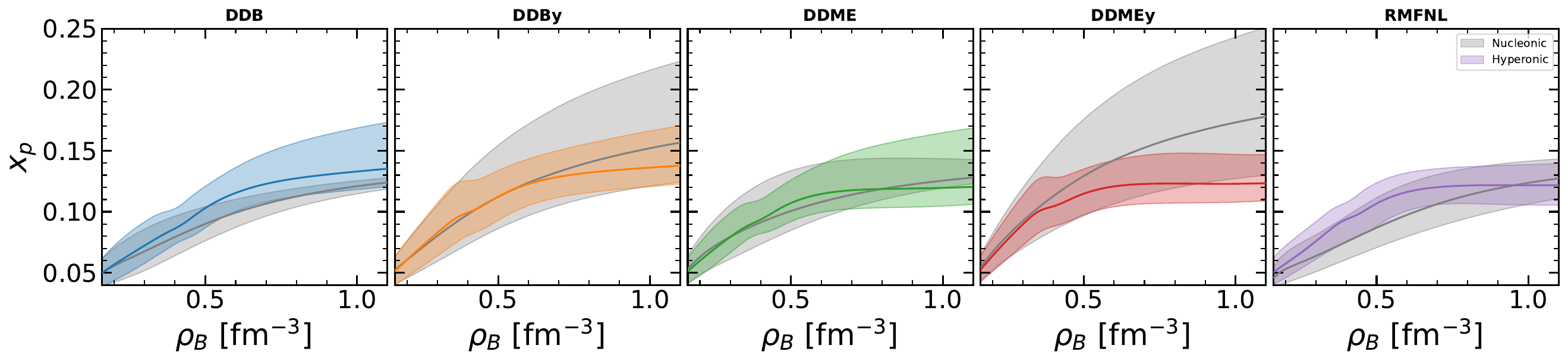}
\caption{ Proton fraction $x_p$ vs.\ baryon density $\rho_B$ for all five model families. Gray bands: nucleonic 90\% CI; colored bands: hyperonic 90\% CI. Solid lines: median. In the $y$-models (DDBy, DDMEy) the nucleonic $x_p$ band is broad because the free $y$ parameter allows a wide range of isovector coupling strengths; the hyperonic posterior constrains $y$ tightly, suppressing $x_p$. In the non-$y$ models (DDB, DDME, RMFNL) the nucleonic band is narrow, and the EOS stiffening required by the $2\,\Msun$ constraint with hyperons increases the symmetry energy at high density, leading to comparable or slightly increased $x_p$.}
\label{fig:proton_fraction}
\end{figure*}

\begin{figure*}[t]
\centering
\includegraphics[width=0.95\textwidth]{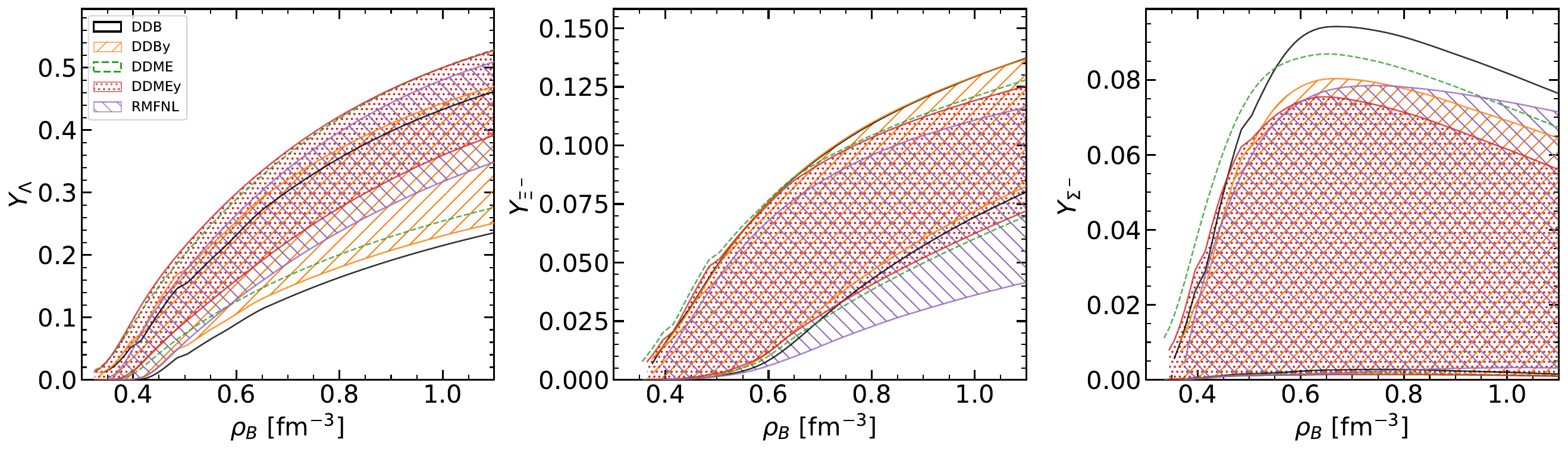}
\caption{Hyperon fractions $Y_\Lam$ (left), $Y_{\Xi^-}$ (center), and $Y_{\Sigma^-}$ (right) vs.\ baryon density for all five hyperonic models, shown as 90\% CI bands. DDB: black solid outline; DDBy: orange hatched (//); DDME: green dashed outline; DDMEy: purple dot hatched; RMFNL: cyan backslash hatched.}
\label{fig:hyperon_fractions_3panel}
\end{figure*}

We further compare: i) the proton fraction between nucleonic and hyperonic posteriors, Fig.~\ref{fig:proton_fraction}; ii) the $\Lambda$, $\Xi^-$ and $\Sigma^-$ fractions across the five models, Fig.~\ref{fig:hyperon_fractions_3panel}. In both cases, the bands denote the 90\% credible regions.

First, we discuss the proton fraction $x_p$ and compare the nucleonic and hyperonic posteriors for each model family. The response of $x_p$ to the inclusion of hyperons depends critically on the flexibility of the isovector channel, controlled by the $\varrho$-meson coupling $h_\varrho(x)$.

In the $y$-models (DDBy, DDMEy) the coupling $h_\varrho(x) = y\exp[-a_\varrho(x-1)] + (1-y)$, Eq.~(\ref{eq:hy}), interpolates between a pure exponential softening ($y=1$) and a density-independent constant ($y=0$). Small $y$ values keep the isovector coupling nearly constant at high density ($h_\varrho \to 1-y \approx 1$), maintaining a large symmetry energy and therefore yielding high proton fractions; large $y$ values cause $h_\varrho$ to drop, reducing the symmetry energy and $x_p$. In the nucleonic posterior $y$ is essentially unconstrained: $y = 0.58^{+0.37}_{-0.31}$ (DDBy), $0.49^{+0.43}_{-0.32}$ (DDMEy) at 90\% CI. This freedom produces a broad $x_p$ distribution whose 90\% CI width at $\rho_B = 0.8$~fm$^{-3}$ reaches $\Delta x_p \simeq 0.09$ (DDBy) and $0.10$ (DDMEy), with upper bounds as high as $x_p \sim 0.20$--$0.22$.

When hyperons are included, the $y$ posterior collapses to $y \simeq 0.47 \pm 0.04$ in both models: the hyperon--meson couplings, including the isovector channel, control the amount of negatively charged hyperons $\Sigma^-$ and $\Xi^-$. At the same time, electrons are suppressed. As a result, the fraction of protons is controlled by charge neutrality, which is now reached with massive negatively charged baryons. The narrowing of the $y$ posterior eliminates the low-$y$ tail that produced the high-$x_p$ solutions.
The net effect is a clear suppression of the proton fraction, with the median shifting by $\Delta x_p^{\rm med} \simeq -0.04$ (DDMEy) and $-0.01$ (DDBy) at $\rho_B = 0.8$~fm$^{-3}$.

For the models without $y$ (DDB, DDME, RMFNL), the $\varrho$-meson coupling is constant (RMFNL) or follows a fixed functional form --- pure exponential (DDB) or rational (DDME) --- with no free mixing parameter. The nucleonic $x_p$ is consequently narrow ($\Delta x_p \lesssim 0.04$ at $0.8$~fm$^{-3}$). When hyperons are added, the system has more freedom to attain electric neutrality. In addition, the $2\,\Msun$ constraint requires a stiffer EOS to compensate for the softening from additional baryonic degrees of freedom. The posterior shifts to higher incompressibility ($\Delta K_0 \simeq +14$ to $+39$~MeV) and less negative symmetry energy curvature ($\Delta K_{\rm sym} \simeq +14$ to $+46$~MeV), while $L_{\rm sym}$ changes only modestly ($\lesssim 3$~MeV for DDB and RMFNL). The combined effect of the higher-order symmetry energy terms enhances the symmetry energy at supra-saturation density, which \emph{increases} $x_p$ in beta equilibrium. This stiffening-driven increase competes with the charge-neutrality reduction from $\Sigma^-$ and $\Xi^-$. In DDB and RMFNL the stiffening dominates ($\Delta x_p^{\rm med} \simeq +0.01$ to $+0.02$), while in DDME the two effects largely cancel ($\Delta x_p^{\rm med} \approx 0$), because $L_{\rm sym}$ decreases by $\sim 6$~MeV partially offsetting the $K_{\rm sym}$ increase of $+37$~MeV.

The fraction of protons is smaller for the RMFNL model because it predicts the smallest amount of negatively charged hyperons. On the other hand, models DDB and DDBy predict the highest amount of negatively charged hyperons at high densities, which explains the larger proton fractions.

\subsection{Hyperon onset densities}

\begin{figure*}
    \centering
    \includegraphics[width=0.99\linewidth]{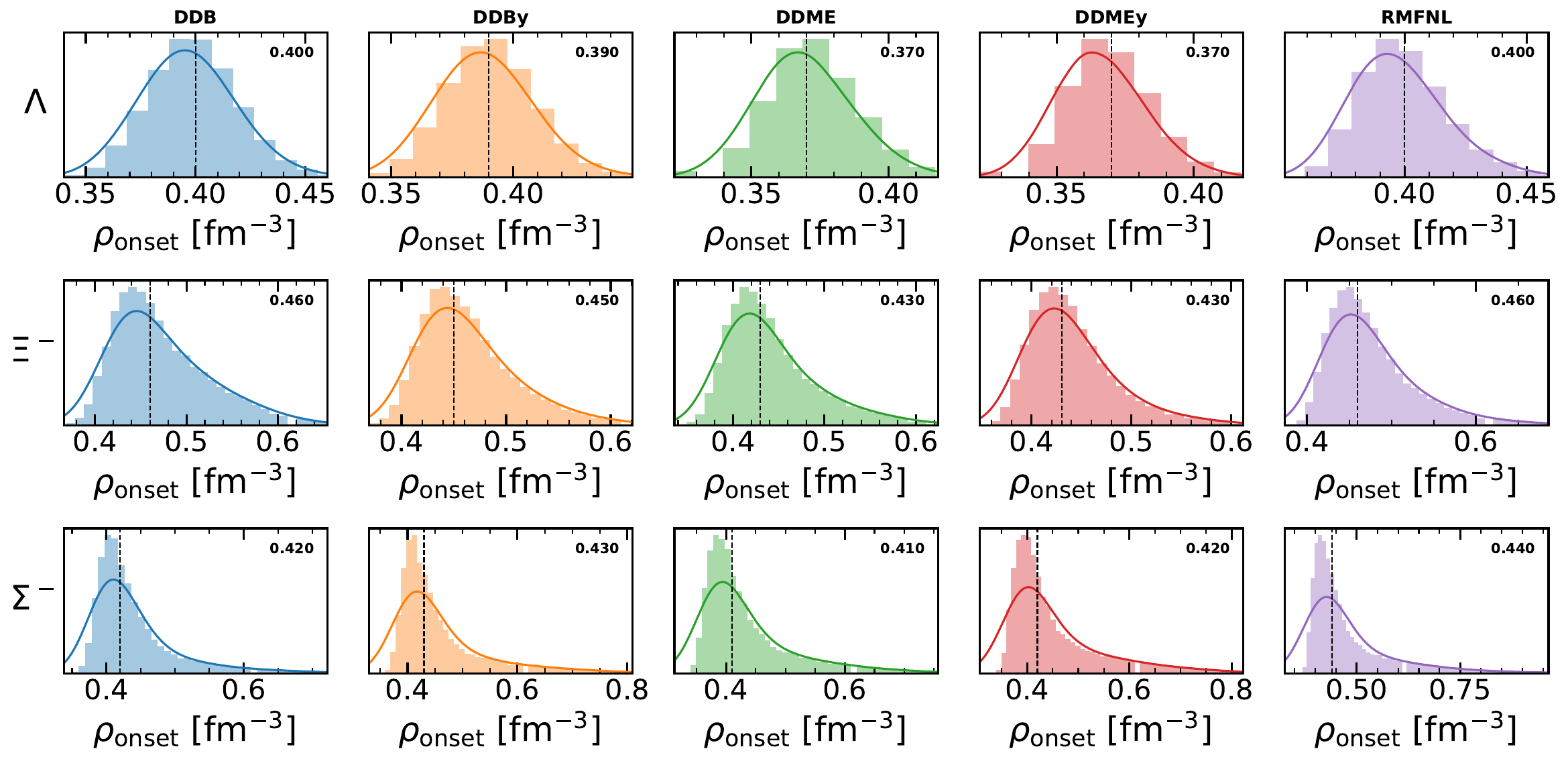}
    \caption{ Posterior distributions of hyperon onset densities in $\beta$-equilibrium neutron star matter for all five hyperonic models. Columns          correspond to the five RMF parametrizations (DDB, DDBy, DDME, DDMEy, RMFNL), while rows show the onset density for $\Lambda$ (top), $\Xi^-$ (middle),  and $\Sigma^-$ (bottom). The onset density is defined as the baryon density at which the particle fraction first exceeds $10^{-4}$, determined by     linear interpolation between successive density grid points. Histograms show the raw posterior distributions and solid curves represent kernel density  estimates. Dashed vertical lines and numerical annotations indicate the median onset density for each model. The $\Lambda$ hyperon appears first in all models, with a median onset at $\rho \approx 0.36$--$0.39$~fm$^{-3}$ ($\approx 2.3$--$2.5\,n_0$), followed by $\Sigma^-$ at $\rho \approx  0.40$--$0.43$~fm$^{-3}$ and $\Xi^-$ at $\rho \approx0.42$--$0.45$~fm$^{-3}$.}
  \label{fig:hyperon_onset}

\end{figure*}

Figure~\ref{fig:hyperon_onset} shows the posterior distributions of onset densities for $\Lam$, $\Sigma^-$, and $\Xi^-$, for all five hyperonic models. Columns  correspond to the five RMF parametrizations (DDB, DDBy, DDME, DDMEy, RMFNL), while rows show the onset density for $\Lambda$ (top), $\Xi^-$ (middle),  and $\Sigma^-$ (bottom).

In each panel, the median is annotated. The $\Lam$ onset is tightly constrained at $0.35$--$0.43$~fm$^{-3}$ ($\sim 2.3$--$2.9\,\rhosat$), with models DDME and DDMEy predicting an earlier onset than the other models. These two models also predict an earlier onset of the $\Xi^-$ hyperons. Concerning the $\Sigma^-$, all models behave in a similar way. We also confirm that the onset of the $\Sigma^-$ and $\Xi^-$ occurs in a similar range of densities: while the $\Xi^-$ has an attractive interaction in nuclear matter, the $\Sigma^-$ hyperon has a smaller mass. This competition occurs at zero temperature: at finite temperature, the mass value has the strongest effect, and the $\Sigma^-$-hyperon becomes more abundant \cite{Oertel:2016xsn}. Notice that while $\Lambda$ and $\Xi^-$ appear in 100\% of posterior samples,  $\Sigma^-$ is absent in $\sim$3--18\% of samples depending on the model. 

\subsection{Hyperon optical potentials in symmetric nuclear matter}
A key observable that connects the hyperon coupling ratios to experimentally accessible quantities is the single-particle potential (optical potential) of each hyperon species in symmetric nuclear matter at saturation density:
\begin{equation}
U_{Y_i}(n_0) = x_{\omega i}\!\left(\Gamma_\omega + \frac{\partial \Gamma_\omega}{\partial\rho}\,n_0\right)\!\omega_0
- x_{\sigma i}\!\left(\Gamma_\sigma + \frac{\partial \Gamma_\sigma}{\partial\rho}\,n_0^s\right)\!\sigma_0,
\label{eq:optical}
\end{equation}
where $\sigma_0$ and $\omega_0$ are the meson mean fields in symmetric nuclear matter at $\rho = n_0$, and $n_0^s$ is the corresponding scalar density. For the RMFNL model with constant couplings, the derivative terms vanish:
\begin{equation}
U_{Y_i}(n_0) = x_{\omega i}\,g_\omega \,\omega_0
- x_{\sigma i}\,g_\sigma \,\sigma_0,
\label{eq:opticalNL}
\end{equation} The optical potential $U_Y$ encodes the net attraction or repulsion experienced by a hyperon $Y$ immersed in nuclear matter at saturation and is directly constrained by hypernuclear experiments.

\begin{figure*}[t]
\centering
\includegraphics[width=0.95\textwidth]{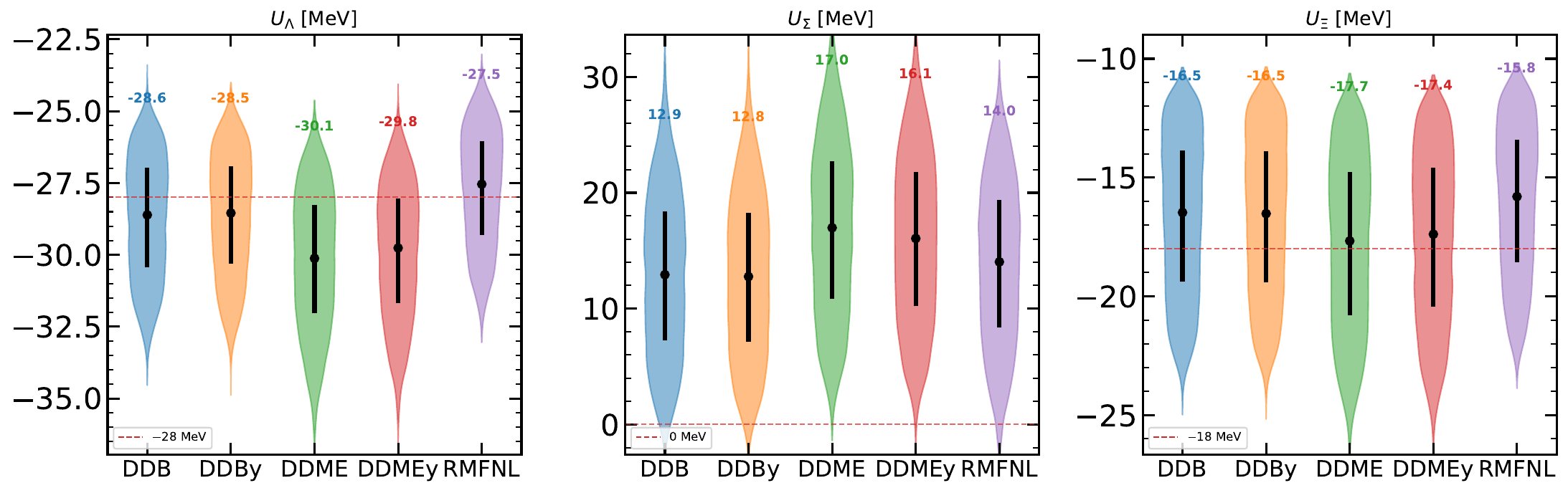}
\caption{Posterior distributions of the hyperon optical potentials $U_\Lam$ (left), $U_\Sigma$ (centre), and $U_\Xi$ (right) in symmetric nuclear matter at saturation density. All five hyperonic models overlaid. Text annotations show the median$^{+1\sigma}_{-1\sigma}$ in MeV.}
\label{fig:optical}
\end{figure*}

Figure~\ref{fig:optical} shows the posterior distributions of $U_\Lam$, $U_\Sigma$, and $U_\Xi$ for all five hyperonic models, computed from the full posteriors by solving symmetric nuclear matter at $\rho_0$ for each parameter sample.

The $\Lam$ potential is attractive across all models, with median values in the range $U_\Lam \approx -28$ to $-34$~MeV, consistent with the experimentally determined value of $U_\Lam \approx -30$~MeV from $\Lam$ hypernuclei~\cite{Gal:2016boi}. The DDH models (DDB, DDBy, DDME, DDMEy) yield $U_\Lam \approx -28$ to $-30$~MeV with uncertainties of $\pm 2.5$~MeV, while the RMFNL model predicts a somewhat deeper potential of $U_\Lam \approx -34$~MeV. This difference arises from the different density dependence in the NL model.

The $\Sigma$ potential is repulsive in all models, with $U_\Sigma \approx +6$ to $+17$~MeV, consistent with the analysis of $\Sigma^-$ atoms which suggests a repulsive potential of $U_\Sigma \approx +10$ to $+50$~MeV~\cite{Friedman:2007zza}. The broad distributions ($\sim \pm 8$~MeV) reflect the relatively wide prior on $x_{s\Sigma} \in [0.43, 0.51]$. The RMFNL model gives the weakest repulsion ($U_\Sigma \approx +6$~MeV), while DDME shows the strongest ($U_\Sigma \approx +17$~MeV).

The $\Xi$ potential is moderately attractive, $U_\Xi \approx -16$ to $-19$~MeV, with the DDH models clustered around $-17$~MeV and RMFNL at $-19$~MeV. These values are consistent with the analysis of $\Xi^-$ production in heavy-ion collisions and recent lattice QCD results, which suggest $U_\Xi \approx -14$ to $-20$~MeV at saturation~\cite{Hiyama2020}.

The consistency of the posterior optical potentials with experimental constraints provides important validation that the sampled coupling ratios ($x_{s\Lam}, x_{s\Sigma}, x_{s\Xi}$) produce physically reasonable hyperon interactions.

\subsection{Slope of the mass-radius curves}
\begin{figure*}
    \centering
    \includegraphics[width=0.99\linewidth]{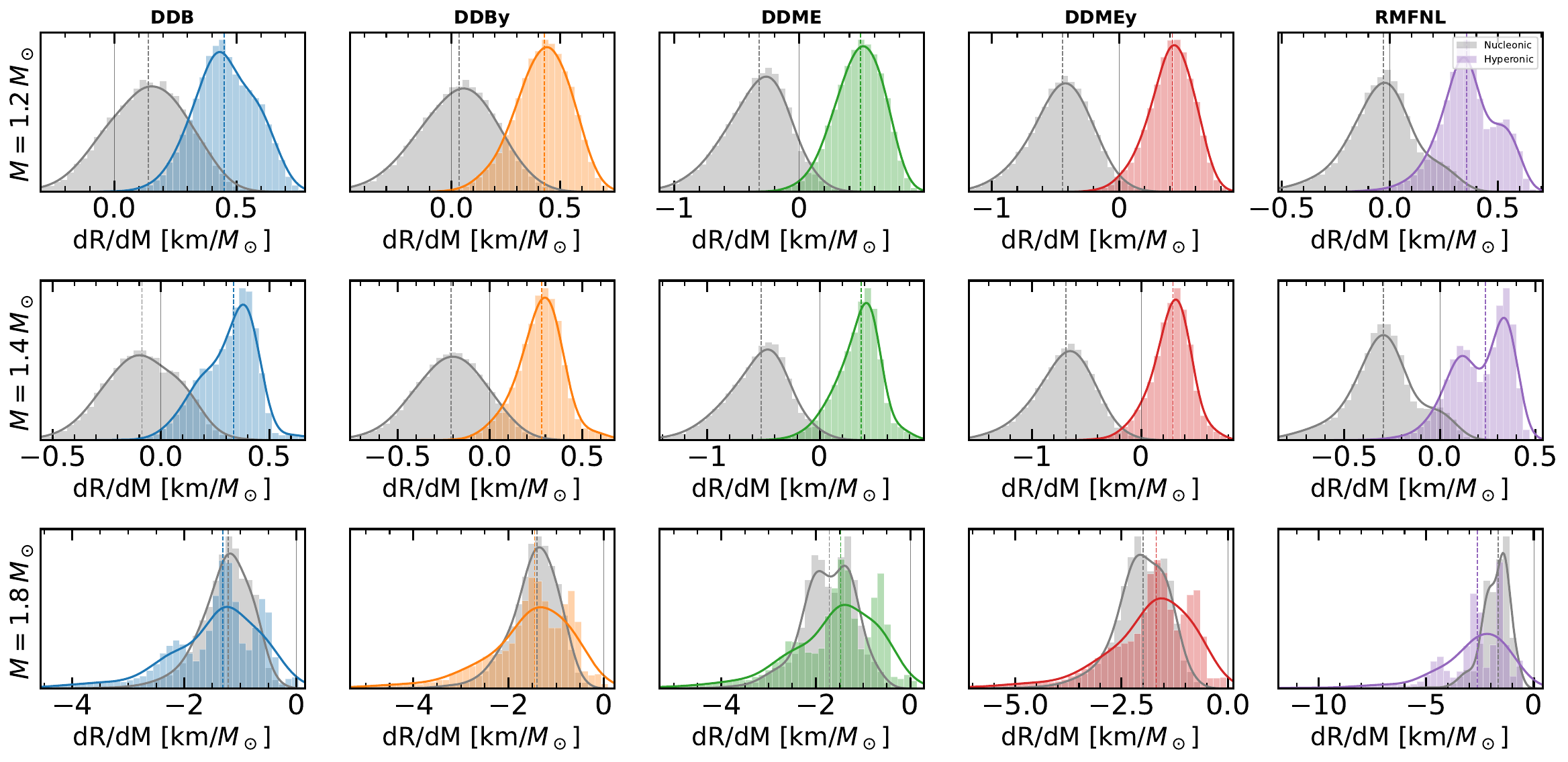}
    \caption{Posterior distributions of the mass-radius slope $dR/dM$ (in km/$M_\odot$) evaluated at three fixed neutron star masses: $M = 1.2\,M_\odot$   
  (top), $1.4\,M_\odot$ (middle), and $1.8\,M_\odot$ (bottom), for all five RMF parametrizations. The slope is computed as a finite difference $\Delta R 
  / \Delta M$ on a uniform mass grid ($\Delta M = 0.01\,M_\odot$) obtained by cubic-spline interpolation of the TOV mass-radius sequences. Gray          
  histograms and curves show the nucleonic posterior, while colored histograms and curves show the corresponding hyperonic posterior. Dashed vertical    
  lines indicate the medians. At $M = 1.2$ and $1.4\,M_\odot$, the nucleonic models yield $dR/dM \lesssim 0$ (the radius has already begun to decrease
  with increasing mass), whereas the hyperonic posteriors consistently favour $dR/dM > 0$ (the radius is still increasing). This sign difference reflects
   the distinct nuclear matter parameter space selected by the astrophysical constraints when hyperons are included: the hyperonic posteriors prefer
  stiffer low-density EOS to compensate for the softening at higher densities where hyperons appear. At $M = 1.8\,M_\odot$ both cases show $dR/dM < 0$,
  with the hyperonic distributions shifted to more negative values, consistent with the softer high-density EOS due to hyperon degrees of freedom.}
  \label{fig:dRdM}

\end{figure*}

It has been shown that the slope of the mass-radius may give some indication about the presence of non-nucleonic matter inside NS \cite{Ferreira:2025dat,Albino:2025puc,Ferreira:2024hxc}. In particular, it was shown that the presence of hyperons \cite{Ferreira:2025dat} and quarks \cite{Albino:2025puc} inside NS that describe two solar mass stars will present with a very large probability a positive slope of the mass-radius curves for masses below $\sim 1.5\, M_\odot$. Moreover, a positive slope at 1.8 solar masses would be a very strong indication of the presence of non-nucleonic degrees of freedom inside NS. 

The radii of different mass NS given in Table \ref{tab:radii}, and discussed in Sec. \ref{sec:eos_mr_cs2} have already confirmed this trend. To complete the information given in  Table \ref{tab:radii}, we plot in Fig. \ref{fig:dRdM}, the slope $dR/dM$ distributions for the five models with and without hyperons, for 1.2, 1.4 and 1.8\, $M_\odot$.  The following conclusions are in order:  i) only a residual number of hyperonic EOS give rise to a  mass-radius curve with a negative slope at   1.2$M_\odot$; ii) for 1.4$M_\odot$ NS, the trend is similar, although the number of EOS with a negative slope increases slightly; iii)  most of the  nucleonic EOS give rise to a negative slope at  1.2$M_\odot$, except for DDB and DDBy. However, at 1.4$M_\odot$ the number of nucleonic EOS with a positive slope is residual for DDME and DDMEy, and  small for RMFNL and DDBy; 
iv) at 1.8$M_\odot$ only the hyperonic EOS may give rise to a mass-radius curve with a positive slope.

\subsection{Nuclear matter  properties}

Tables~\ref{tab:nmp} summarises the posterior distributions. The NMP values are nearly identical between nucleonic and hyperonic versions, except for $K_0$ and $K_{\rm sym}$. The incompressibility generally takes a larger value for the hyperonic EOS (except for DDME), and the difference can be as large as $\sim 40$~MeV. This reflects the fact that, in order to describe two solar mass stars, the EOS including hyperons has to be stiffer to compensate for the softening when hyperons set in. Concerning $K_{\rm sym}$, it systematically shows a less negative value for the hyperonic stars, reaching values close to zero or even positive at 90\% CI. Since no conditions have been imposed that really constrain composition, such as the proton fraction, the values $K_{\rm sym}$ takes are probably just contributing to the stiffening of the EOS, in addition to $K_0$.

\begin{table*}[t]
\caption{Median values and 90\% CI for nuclear matter properties at saturation density.}
\label{tab:nmp}
\begin{ruledtabular}
\begin{tabular}{lccccccccccccccc}
& \multicolumn{3}{c}{$e_0$ [MeV]} & \multicolumn{3}{c}{$K_0$ [MeV]} & \multicolumn{3}{c}{$J_{\rm sym}$ [MeV]} & \multicolumn{3}{c}{$L_{\rm sym}$ [MeV]} & \multicolumn{3}{c}{$K_{\rm sym}$ [MeV]} \\
\cmidrule(lr){2-4} \cmidrule(lr){5-7} \cmidrule(lr){8-10} \cmidrule(lr){11-13} \cmidrule(lr){14-16}
Model & Med. & Min & Max & Med. & Min & Max & Med. & Min & Max & Med. & Min & Max & Med. & Min & Max \\
\midrule
DDB       & $-16.00$ & $-16.32$ & $-15.69$ & 224 & 197 & 269 & 31.5 & 28.7 & 34.1 & 42 & 30 & 56 & $-108$ & $-140$ & $-61$ \\
DDB+Hyp   & $-16.00$ & $-16.31$ & $-15.69$ & 257 & 228 & 296 & 31.3 & 28.6 & 33.9 & 45 & 34 & 60 & $-94$ & $-121$ & $-53$ \\[2pt]
DDBy      & $-16.00$ & $-16.32$ & $-15.69$ & 217 & 191 & 261 & 31.9 & 29.1 & 34.5 & 49 & 35 & 64 & $-66$ & $-119$ & $-8$ \\
DDBy+Hyp  & $-16.00$ & $-16.31$ & $-15.68$ & 257 & 228 & 294 & 31.6 & 28.9 & 34.2 & 51 & 38 & 65 & $-61$ & $-106$ & $0$ \\[2pt]
DDME     & $-16.00$ & $-16.31$ & $-15.68$ & 245 & 197 & 286 & 32.0 & 29.3 & 34.5 & 56 & 42 & 74 & $-116$ & $-136$ & $-78$ \\
DDME+Hyp & $-15.99$ & $-16.29$ & $-15.69$ & 238 & 192 & 297 & 31.6 & 29.1 & 34.1 & 50 & 37 & 67 & $-79$ & $-106$ & $-19$ \\[2pt]
DDMEy    & $-15.99$ & $-16.29$ & $-15.68$ & 244 & 201 & 286 & 32.3 & 29.7 & 34.7 & 62 & 48 & 80 & $-79$ & $-120$ & $-27$ \\
DDMEy+Hyp& $-15.99$ & $-16.29$ & $-15.69$ & 246 & 202 & 305 & 31.8 & 29.4 & 34.3 & 57 & 43 & 74 & $-53$ & $-93$ & $5$ \\[2pt]
RMFNL     & $-16.00$ & $-16.30$ & $-15.70$ & 252 & 213 & 292 & 31.4 & 28.8 & 34.0 & 43 & 32 & 57 & $-148$ & $-188$ & $-70$ \\
RMFNL+Hyp & $-16.00$ & $-16.30$ & $-15.69$ & 266 & 234 & 315 & 31.1 & 28.6 & 33.8 & 45 & 36 & 58 & $-102$ & $-155$ & $-34$ \\
\end{tabular}
\end{ruledtabular}
\end{table*}

\section{Discussion and Conclusions}
\label{sec:conclusions}

We have performed a systematic Bayesian comparison of hyperonic NS matter across five RMF model families, all constrained by the same observational and theoretical data. Our main findings:

\begin{enumerate}
\item \textbf{Hyperon puzzle alleviated:} All models produce
      $M_{\rm max} \gtrsim 1.9\,\Msun$ (90\% CI). The repulsive $\phi$-meson interaction plays an important role in sustaining massive hyperonic stars.

\item \textbf{Radius increase:} $R_{1.4}$ increases by $0.5$--$0.8$~km
      with hyperons, due to the stiffer low-density EOS required to
      support $2\,\Msun$ when the high-density EOS is softened.

\item \textbf{Model dependence:} RMFNL shows the strongest effect
      ($\Delta M_{\rm max} \approx -0.10\,\Msun$); DDH models show
      $-0.05$ to $-0.09\,\Msun$.

\item \textbf{$c_s^2$ signature:} Universal softening at
      $2$--$3\,\rhosat$ in all hyperonic models.

\item \textbf{NMP stability:} Saturation properties are practically insensitive
      to hyperons, except for $K_0$ and $K_{\rm sym}$ that increase if hyperons are included in order to stiffen the EOS.

\item \textbf{Mass-radius slope:} A positive slope of the mass-radius curve at low mass could be an indication of the presence of hyperons inside the NS. A positive slope at $1.8\,M_\odot$ indicates that non-nucleonic degrees of freedom should be present inside the star.

\item \textbf{Proton fraction:} The response of $x_p$ to hyperons depends on the flexibility of the isovector channel. In models with the $y$ parameter (DDBy, DDMEy), the  proton fraction spans a wide range that is narrowed by hyperonic constraints, resulting in net suppression. In models without $y$ (DDB, DDME, RMFNL), the EOS stiffening required to satisfy the $2\,\Msun$ constraint enhances the symmetry energy at high density, leading to comparable or slightly increased $x_p$, compared to nucleonic stars. In summary, the extra flexibility of nucleonic models is visibly suppressed by hyperons, meaning that the average proton distribution is independent of model flexibility when hyperons are included, i.e. that the proton fraction is determined by the hyperon content.
\end{enumerate}

\begin{acknowledgments}
This work was partially supported by national funds from FCT (Fundação para a Ciência e a Tecnologia, I.P, Portugal) under project UID/04564/2025, identified by DOI 10.54499/UIDB/04564/2025, and project  2024.16290.PEX identified by DOI  identifier 10.54499/2024.16290.PEX. P.S would like to acknowledge his PhD grant 2025.01048.BD.  The computational component of this research was carried out on the
Deucalion HPC platform in Portugal within the EURO HPC project EHPC-REG-2025R01-021, whose resources and technical assistance are gratefully acknowledged. 
\end{acknowledgments}

\bibliography{biblio}
\end{document}